\documentclass[]{aa}  

\usepackage{natbib}
\bibpunct{(}{)}{;}{a}{}{,}
\usepackage{graphicx}
\usepackage{revsymb}	
\usepackage{amsmath}	
\usepackage[usenames]{color}	
\usepackage{epstopdf}	
\usepackage{txfonts}
\usepackage{amssymb}
\usepackage{color}
\usepackage[justification=centering]{caption}
\usepackage{geometry} 
\usepackage[parfill]{parskip} 
\usepackage{amssymb}
\usepackage{slantsc}
\usepackage{array}
\usepackage{url}
\usepackage{amsfonts}
\usepackage[colorlinks=true,linkcolor=black, urlcolor=black, citecolor=black]{hyperref}
\usepackage{hyperref} 
\usepackage{sidecap}
\usepackage{graphicx}
\usepackage{xcolor}
\usepackage{nicefrac}
\usepackage{subfigure}
\usepackage{epsfig}
\colorlet{rouge}{red!70!darkgray}

\begin{document}
\title{Seismic inversion of the solar entropy: A case for improving the Standard Solar Model}
\author{G. Buldgen\inst{1}\and S. J. A. J. Salmon\inst{1}\and A. Noels\inst{1} \and R. Scuflaire\inst{1} \and D. R. Reese\inst{2} M-A. Dupret \inst{1} \and J. Colgan \inst{3}\and C. J. Fontes \inst{3}\and P. Eggenberger \inst{4}\and P. Hakel \inst{3}\and D. P. Kilcrease \inst{3} \and S. Turck-Chièze \inst{5}}
\institute{Institut d’Astrophysique et Géophysique de l’Université de Liège, Allée du 6 août 17, 4000 Liège, Belgium \and LESIA, Observatoire de Paris, PSL Research University, CNRS, Sorbonne Universités, UPMC Univ. Paris 06, Univ. Paris Diderot, Sorbonne Paris Cité, 5 place Jules Janssen, 92195 Meudon Cedex, France \and Los Alamos National Laboratory, Los Alamos, NM 87545, USA \and Observatoire de Genève, Université de Genève, 51 Ch. Des Maillettes, CH$-$1290 Sauverny, Suisse \and Laboratoire AIM, CEA$/$DSM - CNRS - Univ. Paris Diderot - IRFU$/$SAp
Centre de Saclay, 91191 Gif$-$sur$-$Yvette Cedex, France}
\date{May, 2017}
\abstract{The Sun is the most constrained and well-studied of all stars. As a consequence, the physical ingredients entering solar models are used as a reference to study all other stars observed in the Universe. However, our understanding of the solar structure is still imperfect, as illustrated by the current debate on the heavy element abundances in the Sun.}
{We wish to provide additional information on the solar structure by carrying out structural inversions of a new physical quantity, a proxy of the entropy of the solar plasma which properties are very sensitive to the temperature gradient below the convective zone.}
{We use new structural kernels to carry out direct inversions of an entropy proxy of the solar plasma and compare the solar structure to various standard solar models built using various opacity tables and chemical abundances. We also link our results to classical tests commonly found in the literature.}
{Our analysis allows us to probe more efficiently the uncertain regions of the solar models, just below the convective zone, paving the way for new in-depth analyses of the Sun taking into account additional physical uncertainties of solar models beyond the specific question of chemical abundances.}{}
\keywords{Sun: helioseismology -- Sun: oscillations -- Sun: fundamental parameters}
\maketitle
\section{Introduction}
	Hitherto, helioseismology has led to striking successes. The precise location of the base of the convective envelope at a fractional radius of $0.713\pm 0.001$ \citep{KosConv}, the inversion of the solar sound speed, density and rotation profiles \citep{Antia94,SchouRot}, the determination of the helium mass fraction in the convective envelope at $Y=0.2485 \pm 0.0035$ \citep{BasuYSun}, and the outcome of the ``solar neutrino problem'' \citep{BahcallNeutrino} are amongst the greatest achievements in this field. In the $90s$, the internal structure of the Sun was extremely well reproduced by Standard Solar Models (hereafter SSMs). Therefore, the physical ingredients of these numerical models, particularly the solar chemical element abundances \citep[][hereafter GN93]{GrevNoels}, were applied to stars other than the Sun and used to compute grids of stellar models. Such grids are one of the basic components in various fields such as stellar population analysis, Galactic evolution, and exoplanetology for example. 
		
	Later on, the physical ingredients of the solar models, such as the equation of state (\cite{Rogerseos}) or the heavy element abundances were continuously refined \citep[][hereafter GS98]{GreSauv}, but, the changes being quite small, the agreement of the models with helioseismology remained. However, two refinements with stronger impacts were more recently brought forward.
	
	The first one addressed the solar chemical mixture with a revised set of heavy element abundances published by Asplund and collaborators \citep{AGS04O,AGS05C}. The abundant (C,N,O) heavy elements saw a strong decrease of their abundances and the metallicity of the Sun was thus reduced by about $30\%$. Using these new results led to strong disagreements between SSMs and helioseismology \citep{SerenelliComp}. Further revision of the spectroscopic determinations \citep[][hereafter AGSS09]{AGSS09} led to slight reincreases of the metallicity, but were insufficient to restore the agreement with helioseismology. These discrepancies were suggested to originate in additional physical processes acting in the solar radiative zone \citep{KumarWaves,CastroRichard}, but none of these attempts provided a clear and decisive answer to the issue. Simultaneously, other studies used seismology to estimate the solar metallicity. Some confirmed the GS98 values \citep{BasuAbundProc} while others agreed with the AGSS09 values \citep{VorontsovSolarEnv2014}, illustrating the stalemate of this problem. 
	
	The second important change was the revision of the stellar material opacity. The solar problem has been linked to the opacity at the base of the convective envelope and a process inducing a local increase of the opacity has recurrently been proposed as the solution to the controversy \citep{BasuReview}. Until recently, the most commonly used opacities were the OPAL opacity tables \citep{OPAL} but an underestimate of the opacity in more massive stars was convincingly revealed by different studies \citep{SalmonOpac,Cugier12,TurckOpac}. These findings initiated both innovative measurements with high-energy laser facilities and numerical computational efforts to improve theoretical calculations. The first experimental results for iron revealed an important discrepancy with theoretical expectations \citep{Bailey}. In parallel, two new sets of theoretical opacities were developed, one dedicated to the Sun from the OPAS consortium \citep{Mondet} and the other covering the wide range of stellar conditions, by the Los Alamos National Laboratory \citep[][hereafter OPLIB opacities]{Colgan}, which could become commonly used in stellar models. 
	
	The solar issue impacts astrophysics as a whole since the ``metallicity scale'', used to relate spectroscopic observations to the metallicity of stellar models, takes the Sun as its reference. To this day, the so-called ``solar metallicity problem'' remains a tedious issue which is not only linked to the metallicity, but to the whole micro- and macrophysical representation of the stellar structure. Indeed, asteroseismic results have already shown that our depiction of transport processes in stellar models is imperfect \citep[e.g.][]{MosserRot,Deheuvels2014}. Due to the quality of the solar data, the Sun still constitutes our best laboratory to test the ingredients of stellar models. Consequently, providing new seismic diagnostics allowing a more in-depth probe of the solar structure is crucial. With this study, we provide such a new diagnostic by performing structural inversions of an entropy proxy. The sensitivity of this inversion to the stratification just below the convective zone paves the way for a re-analysis of the importance of additional physical processes required in the description of the solar structure. In the following section, we show how our diagnostic sheds new light on the solar structure problem.   

\section{Inversion of the solar entropy: a new seismic diagnostic}\label{SecSInv}

\subsection{Inversion for Standard Solar Models}
	The models considered in this study are SSMs, built with the Liège stellar evolution code \citep[CLES,][]{ScuflaireCles}. The frequencies were computed with the Liège oscillation code  \cite[LOSC,][]{ScuflaireOsc}. All models presented in this paper are computed using the Free equation of state \citep{Irwin} and either the OPAL or OPLIB opacity tables. In order to fully estimate the effects of a change in the heavy element abundances, we adopted two extreme mixtures, namely GN93 and AGSS09. The structural kernels and the inversions were computed with an adapted version of the InversionKit software \citep{ReeseDens} using the SOLA technique \citep{Pijpers}. We used the same solar seismic dataset as in \citep{BasuSun} and followed their definitions of the error bars for the inversion. We followed \cite{RabelloParam} to calibrate the free parameters of the SOLA technique and deal with the surface effects contributions.
	 
	Results of sound speed inversions for the new OPLIB opacities are shown in \cite{GuzikLanl} and illustrate that SSMs built using the AGSS$09$ abundances display a slightly deeper convective envelope and slightly better agreement. These improvements are however mitigated by a larger discrepancy with the helium abundance in the convective envelope found at $0.23$. This reduction is also observed for GN$93$ models which now display a value of $0.24$. This trend results from the lower values of the OPLIB opacities in most of the radiative region, which leads to calibrated SSMs with lower initial helium abundances. The improvements of the sound speed profile and the position of the base of the convective envelope do not result from an overall increase of the opacity, but from a steepening of its derivatives which in turn leads to a steepening of the temperature gradient below the convective zone. 
	
\begin{figure*}[h]
	\centering
		\includegraphics[width=11.8cm]{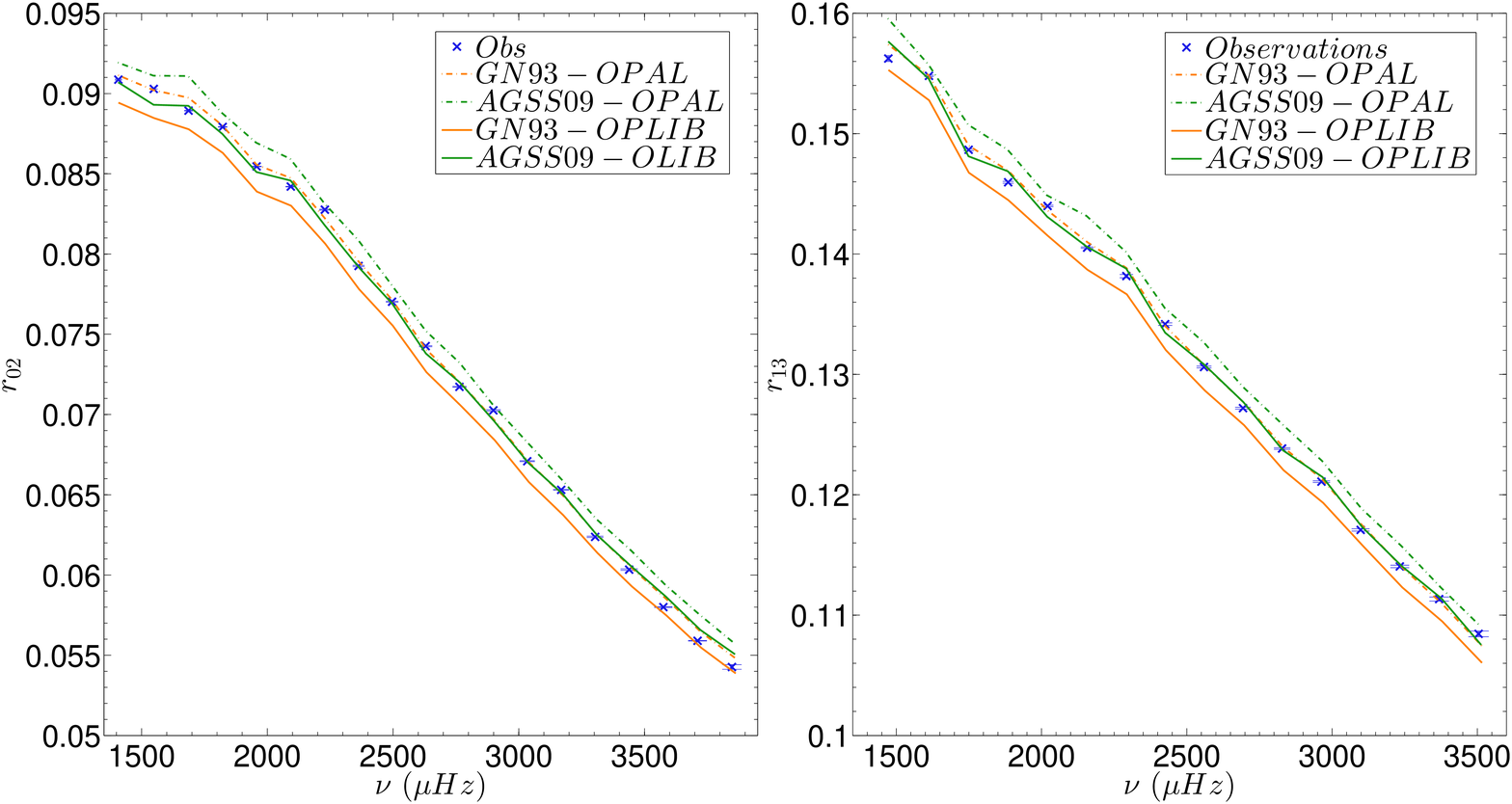}
	\caption{Frequency ratios $r_{02}$ and $r_{13}$ for the Sun and two SSMs. The observations are given with their error bars in blue, while the dashed green line shows the results for an SSM built using the OPAL opacities, the Free equation of state and the AGSS09 abundance tables. The dashed orange line shows the values of these ratios for an SSM built with GN93 abundances, the OPAL opacities and the Free equation of state. The solid green and orange lines shows the ratio values when using the OPLIB opacities instead of the OPAL opacities for the AGSS09 and GN93 abundances respectively. }
		\label{figRatiospapier}
\end{figure*}

	The issue becomes more intricate when one analyses the ratios of the small frequency separation to the large frequency separation. These ratios, denoted $r_{02}$ and $r_{13}$, are used to probe the solar core conditions \citep{RoxburghRatios} and show a clear preference for the latest AGSS09 mixture when the OPLIB opacities are used as is illustrated in Fig. \ref{figRatiospapier}. This constitutes a clear change from the previous SSMs with the OPAL opacities, which showed better agreement with higher metallicity abundances, such as the GN93 or GS98 tables \citep{ChaplinRatios}. Hence, the situation is quite confusing since the sound speed inversion seems to favor the GN93 mixture when using the OPLIB opacities while the frequency ratios better agree with low metallicity models using the same opacity tables. It seems fair to admit that no clear solution emerges from classical helioseismic diagnostics.
	To shed new light on the solar problem, we propose a new seismic diagnostic consisting of inverting a solar entropy proxy, defined as $S_{5/3}=\frac{P}{\rho^{5/3}}$, with $P$, the pressure and $\rho$, the density, which reproduces the behaviour of the entropy of the solar plasma. The kernels used are thus those of the $(S_{5/3},\Gamma_{1})$ pair, with $\Gamma_{1}=\left( \frac{\partial \ln P}{\partial \ln \rho}\right)_{S}$, the adiabatic exponent. The constraining nature of this proxy originates in the plateau that it forms in convective regions. This plateau is due to the high efficiency of convection in the deep layers of the solar envelope, where this phenomenon operates adiabatically. In turn, the height of the plateau is a direct marker of the way we model the radiative zones of the Sun. In the layers below the convective envelope, the stratification is very sensitive to both opacity and chemical abundances. Consequently, a change in opacity, whatever its origin, or a variation of the abundances will impact both the temperature and mean molecular weight gradients and thus the height of the plateau in a given solar model. Testing this height through seismic inversions offers a straightforward diagnostic, complementary to that of the sound speed inversions. Moreover, non-standard processes may also change the height of the plateau, making this diagnostic a very sensitive probe of the layers just below the convective envelope, which are precisely the ones where discrepancies are the largest and where the physical hypotheses of the SSMs are the most uncertain. 
		
\begin{figure*}[h]
	\centering
		\includegraphics[width=13.5cm]{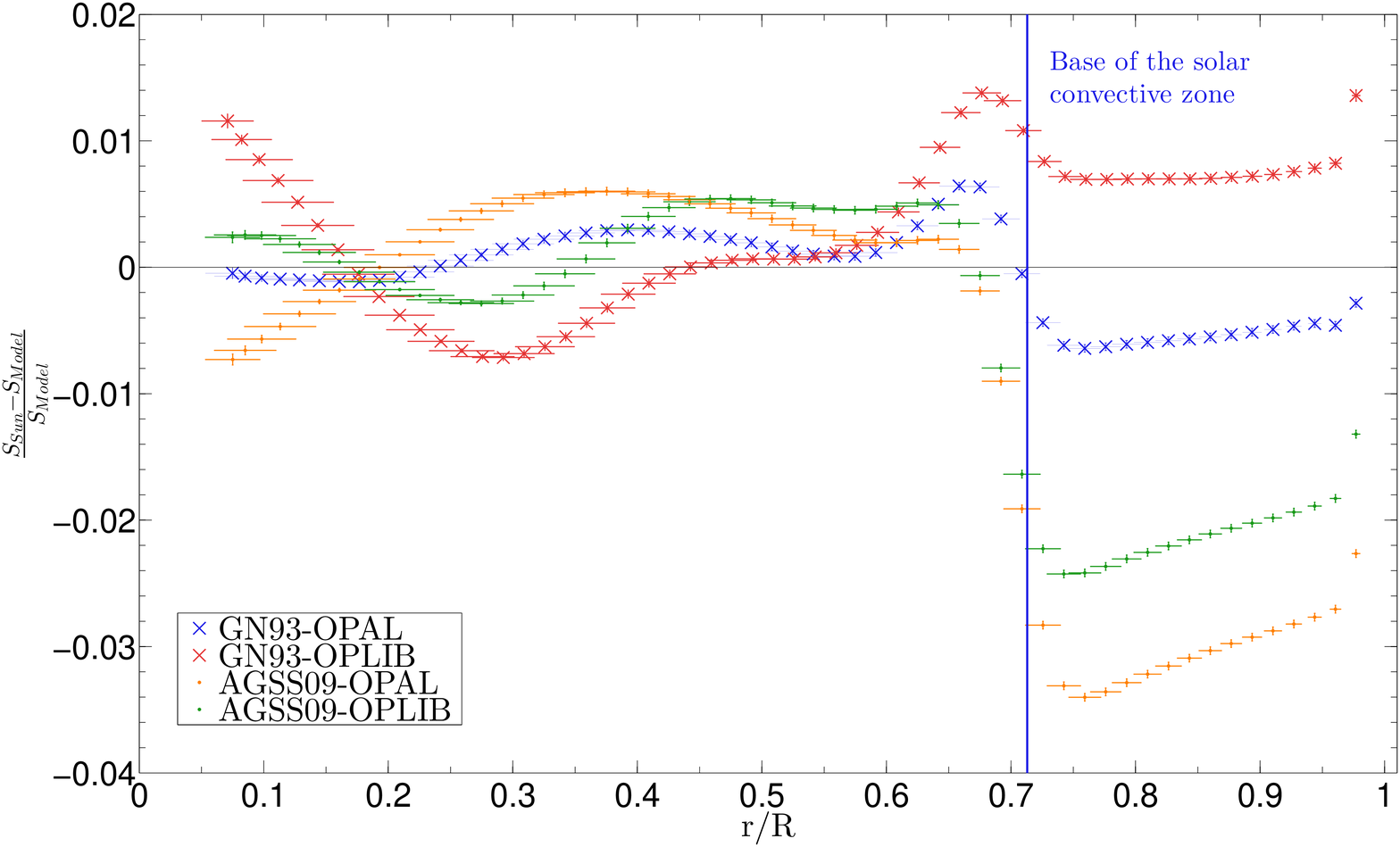}
	\caption{Comparison of the entropy profile between AGSS09 and GN93 SSMs. The red and blue crosses illustrate inversion results for the profile of the entropy proxy in the Sun for the GN93 SSM built with the OPAL and OPLIB opacities. The green and orange circles illustrate the effects of changing from the old OPAL opacities to the recent OPLIB opacities in AGSS09 SSMs.}
		\label{figSpapier}
\end{figure*}

	Inversion results of the entropy proxy profile are given in Fig. \ref{figSpapier} for solar models using either the former OPAL or the new OPLIB opacities. The orange and green circles illustrate the results for the AGSS09 abundances, while the blue and red crosses illustrate the results for the GN93 abundances. We notice that the plateau of the entropy proxy is shifted by about $2\%$ due to the opacity changes between the OPAL and OPLIB opacity tables. While the agreement between the Sun and the GN93 SSMs is still of the order of $0.7\%$, which is quite good, the sign of the differences in the plateau has critical implications. A positive difference between the Sun and the GN93 model built using OPLIB tables means that the entropy plateau in the model is too low. If one were to reconcile the GN93 abundances with the entropy profile of the Sun, it would necessarily require some change inducing a less steep temperature gradient in order to raise the entropy plateau up to the solar value. This appears to be in contradiction with the experimental results of \cite{Bailey} for iron in the physical conditions present at the base of the envelope, which would lead to a strong steepening of the temperature gradient in this region. Furthermore, theoretical calculations of iron spectral opacity in these conditions are still a matter of debate and could as well change in the future \citep{IglesiasIron, Nahar, BlancardComment}.
		
		 In opposition, the models built using the AGSS09 abundances and the OPLIB opacities are in better agreement then the GN93 OPLIB models in most of the radiative region of the solar structure and still show negative differences in the convective envelope. These negative differences mean that a further steepening of the temperature gradient below the convective zone could improve the agreement with the Sun. The entropy inversion being very sensitive to the layers right below the convective zone, it could efficiently constrain non-standard processes. Indeed, additional mechanisms would alter both temperature and mean molecular weight gradients and the changes would be clearly seen in the variations of the height of the plateau.

	In addition to the OPLIB opacities, we have also tested the OPAS opacity tables, which have been optimized for the base of the convective zone \citep{Mondet}. However, these tables do not cover the full solar conditions and are only available for the AGSS$09$ abundances, restricting their potential for comparisons using various physical ingredients. SSMs built using these tables lead to slightly larger discrepancies with the Sun. 
	
	We also tested the dependency of our proxy to the equation of state by using the OPAL $2005$ equation of state instead of the Free equation of state. They induce around ten times smaller differences than those due to the change in opacities or abundances in the models. Therefore, most of the changes in the entropy plateau are to be expected from non-standard processes or updates in the opacity tables.
\subsection{Analysis of the seismic diagnostic of the entropy proxy}
	Our entropy proxy, denoted $S_{5/3}=\frac{P}{\rho^{5/3}}$, comes from the Sackur-Tetrode equation for the entropy of a  mono-atomic non-degenerate ideal gas, which reads  
\begin{align}
S=\frac{3k_{B}}{2}\left(\mu m_{u} \ln \left( \frac{P}{\rho^{5/3}} \right) + f(\mu)\right),
\end{align}
with $k_{B}$ the Boltzmann constant, $\mu$ the mean molecular weight, $m_{u}$ the atomic mass unit, $P$ the local pressure, $\rho$ the density and $f(\mu)$ a function that only depends on the mean molecular weight and physical constants. The most striking advantage of this proxy is its unambiguous behaviour towards opacity changes just below the convection zone. Indeed, if one takes the derivative of the natural logarithm of $S_{5/3}$ with respect to the natural logarithm of $P$ for an ideal gas, one obtains
\begin{align}
\frac{d \ln S_{5/3}}{d \ln P}=\frac{-2}{3}+\frac{5}{3}\left(\frac{d \ln T}{d \ln P} - \frac{d \ln \mu}{d \ln P} \right).
\end{align}
Now, for a given energy flux, an increased opacity below the convection zone induces a steeper temperature gradient against pressure. This, in turn, will increase the logarithmic derivative of $S_{5/3}$ and brings it closer to $0$ since $\frac{d \ln S_{5/3}}{d \ln P}$ is negative. Therefore, the increase in entropy versus the radius is thus smaller as the pressure decreases and the height of the plateau is accordingly reduced with the steepening of the temperature gradient just below the convective envelope. This effect is illustrated in Fig. \ref{figSpapier}, where the near $2\%$ shift results mainly from changing the opacity tables used in the SSMs. The steeper temperature gradient is a consequence of steeper dependence of the OPLIB opacities with temperature \citep[see][]{Colgan}. The effect of a localized ad-hoc opacity increase on this indicator has been observed in all test cases involving the past OPAL and the latest OPLIB and OPAS opacities. These tests on structural models have confirmed the trends we have discussed here.

\subsection{Additional tests of the inversion techniques}
\begin{figure}[t]
	\centering
		\includegraphics[width=7.8cm]{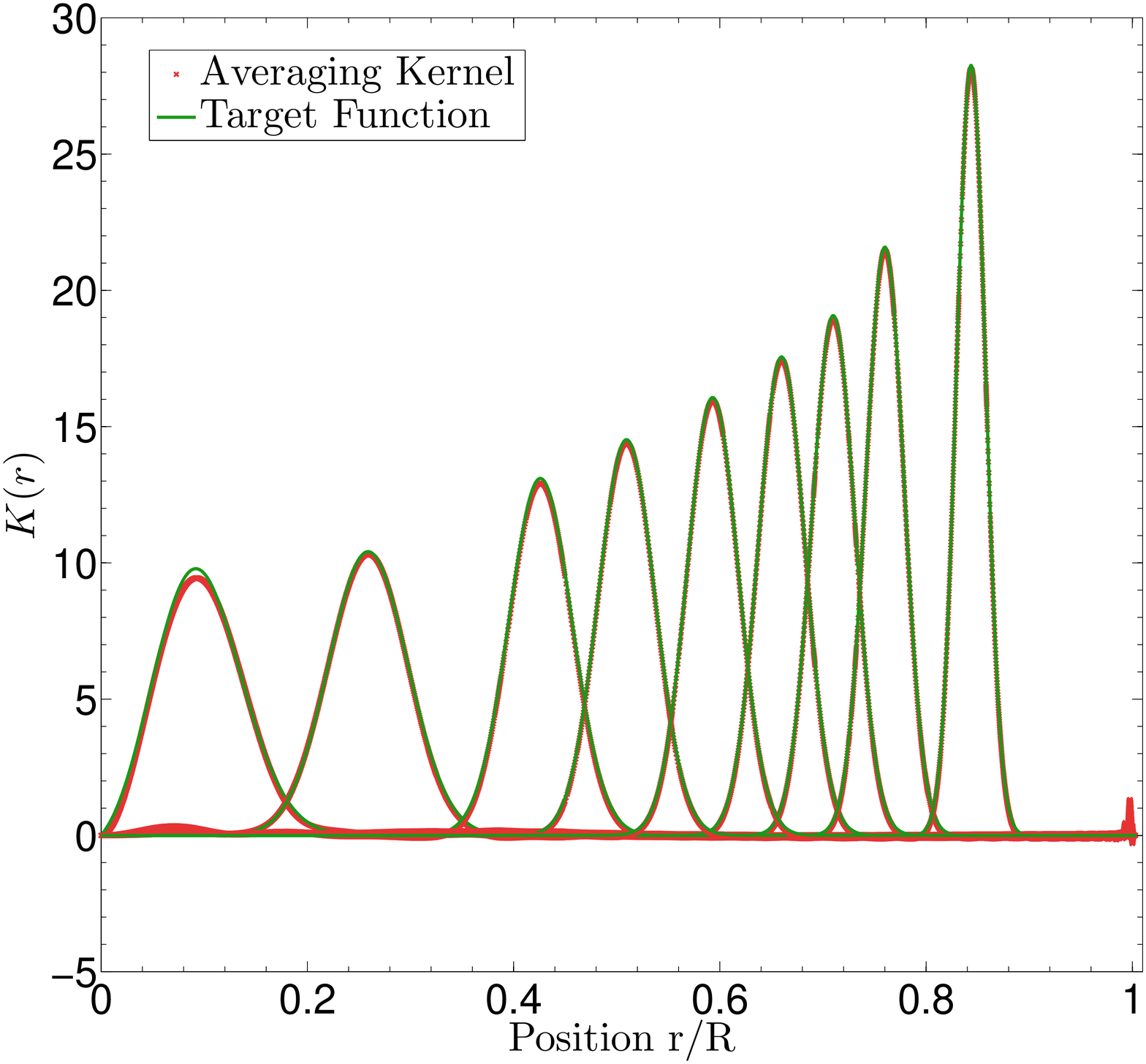}
	\caption{Illustration of SOLA averaging kernels and comparison with their target functions at various depth. The green curves show the Gaussian target functions of the SOLA inversion for various depths inside the Sun while the red dashed curves show the averaging kernels. }
		\label{figKerpapier}
\end{figure}

	 We also performed further checks of the quality of the averaging kernels for the SOLA method. We illustrate in Fig. \ref{figKerpapier} the averaging kernels \citep{Pijpers} of the SOLA method for various positions inside the Sun. One can clearly see that the target function in green is well reproduced at every depth, although some inaccuracies are present below $0.1$ solar radii.  This is expected since we lack very low degree and radial order modes able to probe efficiently the deepest layer of the solar structure.   
\section{Conclusion}
	The change of scenery caused by the use of the OPLIB tables in SSMs points out weaknesses for both high- and low-metallicity abundances tables. An intermediate metallicity value or an increased opacity at the base of the convective envelope could marginally restore the agreement for SSMs, but the discrepancies in helium seem to point out additional mechanisms, some physical ingredients that have to be included in the solar models whatever abundance tables are used. The uncertainties illustrated in this study and the sensitivity of the seismic diagnostic we developed lead us to advocate for a re-opening of the case of potential additional ingredients in helioseismic analyses using constraints such as the lithium abundance and the solar rotational profile in combined studies using simultaneously all seismic information available. Changes in the physical ingredients of solar and stellar models will impact our determinations of stellar fundamental parameters. It is a necessary step if we want to bring these models to a new level of physical accuracy. For that purpose, seismic inversions of the entropy profile offer unprecedented opportunities to further test the structure of the Sun. 
	
\bibliography{biblioarticleS}

\end{document}